\documentclass[aps, amsmath, amssymb, prb,
superscriptaddress, showpacs, preprintnumbers,floatfix]{revtex4}
\usepackage{graphicx}
\usepackage{dcolumn}
\usepackage{bm}
\usepackage{setspace}   

\newcommand{\beq}{\begin{equation}}
\newcommand{\eeq}{\end{equation}}
\newcommand{\ba}{\begin{array}{ccc}}
\newcommand{\ea}{\end{array}}
\newcommand{\nn}{\nonumber}
 \renewcommand{\d}{\partial}
\def\bea{\begin{eqnarray}}
\def\eea{\end{eqnarray}}

\def\Tr{ {\rm Tr} }
\def\<{\langle}
\def\>{\rangle}
\def\cpn{\mathrm{CP}^{N-1}}
\def\cp1{\mathrm{CP}^1}

\begin{document}
\title{Monopoles in $\cpn$ model via the state-operator correspondence}
\author{Max A. Metlitski}
\affiliation{Department of Physics, Harvard University, Cambridge, Massachusetts 02138, USA}
\author{Michael Hermele}
\affiliation{Department of Physics, University of Colorado, Boulder, Colorado 80309, USA}
\author{T. Senthil}
\affiliation{Department of Physics, Massachusetts Institute of Technology, Cambridge, Massachusetts 02139, USA}
\author{Matthew P. A. Fisher}
\affiliation{Microsoft Research, Station Q, University of California, Santa Barbara, California 93106, USA}

\date{\today\\[24pt]}
\vfill
\begin{abstract}
One of the earliest proposed phase transitions beyond the Landau-Ginzburg-Wilson paradigm is the quantum critical point separating an antiferromagnet and a valence-bond-solid on a square lattice. The low energy description of this transition is believed to be given by the $2+1$ dimensional $\cp1$ model -- a theory of bosonic spinons coupled to an abelian gauge field. Monopole defects of the gauge field play a prominent role in the physics of this phase transition. In the present paper, we use the state-operator correspondence of conformal field theory in conjunction with the $1/N$ expansion to study monopole operators at the critical fixed point of the $\cpn$ model. This elegant method reproduces the result for monopole scaling dimension obtained through a direct calculation by Murthy and Sachdev. The technical simplicity of our approach makes it the method of choice when dealing with monopole operators in a conformal field theory. 
\end{abstract}
\vfill
\maketitle

\section{Introduction}

Recent theoretical studies have begun to elucidate
two remarkable classes of quantum critical phenomena in two-dimensional magnetic insulators.  Phase transitions beyond the Landau-Ginzburg-Wilson paradigm make up the first such class.\cite{deconfined1,deconfined2,grover08,ran08}  These Landau-forbidden transitions are continuous quantum critical points (QCPs) between two conventional ordered ground states, where a Landau theory description in terms of the two order parameters does not predict a direct continuous transition upon tuning a single parameter.  The second class consists of critical spin liquids, which are disordered ground states with gapless excitations and power law correlations, and which can exist as stable zero-temperature phases that can be accessed with no fine-tuning of parameters.\cite{affleck88,marston89,lee92,halperin93,polchinski94,rantner01,rantner02,hermele04,sslee08}  Aside from the intrinsic theoretical interest, there is evidence for a Landau-forbidden phase transition in a model of $S = 1/2$ spins, between a Neel antiferromagnet and a valence-bond solid (VBS).\cite{sandvik07,melko08}  Moreover, several materials have emerged as candidates for critical spin liquid ground states.\cite{shimizu03,motrunich05,sslee05,helton07,ofer06,mendels07,ran07,yokamoto07,zhou08,lawler08}

The field-theoretic description of such phenomena can typically be cast in terms of a gauge field coupled to bosonic and/or fermionic matter fields.  In particular, the Landau-forbidden QCP (quantum critical point) between the Neel and VBS ground state is described by the $\cpn$ model for $N=2$,\cite{deconfined1,deconfined2} which consists of an $N$-component boson field $z$ coupled to a \emph{compact} ${\rm U}(1)$ gauge field $A_{\mu}$.
Compactness means that magnetic monopole defects of the gauge field are present and carry the quantized flux $2\pi q$; in two dimensions, these are instanton configurations of the gauge field in space-time.  
Such topological defects, and the field theory operators (called monopole operators) that insert them at a particular point in space-time, play an important role in Neel-VBS transition, and in other gauge theories of Landau-forbidden QCPs and critical spin liquids.  In the present case, $q = 1$ monopole operators play a particularly important role as the order parameter for the VBS state.  Furthermore, $q=4$ monopole operators are allowed perturbations to the action.  Thus it is important to have information about the scaling dimensions of monopole operators, which determine power-law decay of their two-point functions, and whether those operators allowed by symmetry are relevant perturbations to the action.

Many of the gauge theories of interest, including the $\cpn$ model, are solvable in a large-$N$ limit, where the number of bosonic or fermionic matter fields is taken large.  Even in this solvable limit, it is challenging to work with monopole operators, because they cannot be expressed as a polynomial of gauge fields and matter fields.  While electric-magnetic duality gives direct access to monopole operators,\cite{polyakov77} it is limited to purely bosonic theories with only abelian symmetries.  Despite these difficulties, progress has been made:  in a technical \emph{tour de force}, by a direct evaluation of the free energy of a monopole-antimonopole pair, Murthy and Sachdev calculated the monopole scaling dimension as a function of $q$ for the $\cpn$ model in the large-$N$ limit.\cite{murthy90}  Much more recently, Borokhov, Kapustin and Wu exploited the state-operator correspondence of conformal field theory to calculate the monopole scaling dimension for massless Dirac fermions coupled to a ${\rm U}(1)$ gauge field, often referred to as QED3.\cite{borokhov02}  In the large-$N$ limit, calculation of the scaling dimension was reduced to determining the ground state energy of free Dirac fermions moving on a sphere with a background quantized flux.  Although conceptually more sophisticated, this calculation was technically much simpler than that of Murthy and Sachdev. 

In this paper, we follow Ref.~\onlinecite{borokhov02} and apply the state-operator correspondence to calculate monopole scaling dimensions in the $\cpn$ model, and reproduce the result of Murthy and Sachdev in a relatively simple calculation.  In addition to the aesthetic advantage of greater simplicity, this result provides a nontrivial check on the correctness of the Murthy-Sachdev result.  Furthermore, it illustrates the power of the state-operator correspondence in working with monopole operators of conformal field theories in three space-time dimensions. 

The outline of our paper is as follows.  In Sec.~\ref{sec:review} we begin with a brief review of the solution of the  $\cpn$ model in the large-$N$ limit.  Next, in Sec.~\ref{sec:soc} we review the state-operator correspondence in some detail.  In Sec.~\ref{sec:calc}, we use the state-operator correspondence to calculate the monopole scaling dimension in the $\cpn$ model, and present the details of the calculation.  This is followed by a discussion (Sec.~\ref{sec:discussion}) and conclusions (Sec.~\ref{sec:conclusion}).  Technical details are contained in two appendices.

\section{Review of $\cpn$ model}
\label{sec:review}

The Lagrangian of the $\cpn$ model in $D = 3$ Euclidean space-time dimensions is
\begin{equation}
\label{eqn:cpnmodel}
{\cal L} = |D_{\mu} z |^2 + i  \lambda (|z|^2-\frac{1}{g}) \text{,}
\end{equation}
where $z$ is an $N$-component complex scalar field, and  $\lambda$ is a local Lagrange multiplier enforcing the constraint $z^{\dagger} z = 1/g$.  The covariant derivative $D_{\mu} \equiv \partial_{\mu} - i A_{\mu}$, where $A_{\mu}$ is a \emph{non-compact} ${\rm U}(1)$ gauge field.  The non-compactness of $A_{\mu}$ is equivalent to the fact that the gauge flux is a conserved ${\rm U}(1)$ current
$j^{G}_{\mu} = \epsilon_{\mu \nu \lambda} \partial_{\nu} A_{\lambda}$.  Conservation of $j_{\mu}^G$ is equivalent to the absence of monopole events in space-time, or, in other words, to the absence of monopole operators in the Lagrangian.  For the purposes of this paper, there is no need to consider the more complicated \emph{compact} $\cpn$ model, which can be easily defined on the lattice.  The reason is that monopole operators are irrelevant (in the renormalization group sense) at the large-$N$ critical point of the $\cpn$ model, and so the critical properties will be the same whether we start with a compact or non-compact model. 

The global symmetry is thus $( {\rm SU}(N) / {\rm Z}_N ) \times {\rm U}(1)$, where the ${\rm SU}(N)$ rotates among the $N$ components of $z$, and the ${\rm U}(1)$ is the symmetry associated with flux conservation (\emph{i.e.} conservation of $j_{\mu}^G$).  The quantized flux $q$ of a monopole operator is its charge under the ${\rm U}(1)$.  A useful way to state the difference between the compact and non-compact $\cpn$ models is that non-compact model has ${\rm U}(1)$ flux conservation as an exact microscopic symmetry, while in the compact model this symmetry is not present.  However, at least in the large-$N$ limit, this symmetry emerges at long distances at the critical point, corresponding to the irrelevance of monopole operators.

The critical point of the $\cpn$ model is a continuous transition between an ordered phase where $z$ is condensed (small $g$), and a disordered phase (large $g$) where the only low-energy excitation is the photon of the ${\rm U}(1)$ gauge field. Upon integrating out the $z$-bosons, we obtain the effective action for the fields $A_{\mu}$ and $\lambda$,
\beq
 \label{Snonloc}
 S_{{\rm eff}} = N\, \Tr \ln(-D_{\mu} D_{\mu} + i \lambda) - \frac{1}{g} \int d^D x \,i \lambda \text{.}
\eeq
Taking $g \propto 1/N$, $S_{{\rm eff}}$ is exactly solved by the saddle-point approximation in the large-$N$ limit, and corrections to any desired quantity can be obtained in the $1/N$ expansion.

In the large-$N$ limit, monopoles appear as the solutions to the saddle point equations where $\partial_{\mu} j^{\mu}_G \neq 0$ at a few points in space-time.  For example, the lowest action saddle point with a charge-$q$ monopole at the origin has a gauge field $A^q_{\mu}$, chosen so that
\begin{equation}
\epsilon_{\mu \nu \lambda} \partial_{\nu} A^q_{\lambda} = \frac{q}{2} \frac{x_{\mu}}{x^3} \text{.}
\end{equation}
One then needs to solve the saddle point equations to find the saddle-point value of the Lagrange multiplier field, $\bar{\lambda}_q(x)$.  The corresponding saddle-point action of the monopole is then
\begin{equation}
S_q = N \, \Tr \ln(-(\partial_{\mu} - i A^q_{\mu})(\partial_{\mu} - i A^q_{\mu}) + i \bar{\lambda}_q) - \frac{1}{g} \int d^D x \,i \bar{\lambda}_q \text{.}
\end{equation}
At the critical point ($g = g_c$), the action $S_q$ is related to the scaling dimension of the monopole operator $m^*_q(x)$, which inserts a charge-$q$ monopole.  To see this, we put the theory in a space-time which is a ball of radius $R$.  Then we consider the object
\begin{eqnarray}
f(R) &=&  \langle m^*_q(0) \rangle  \\
&=&
\frac{ \int [d z] [d A_{\mu}] [d \lambda] \, m^*_q(0) e^{- \int_{|x| < R} d^3 x \, {\cal L}  } }{\int [d z] [d A_{\mu}] [d \lambda] e^{- \int_{|x| < R} d^3 x \, {\cal L}  } } \\
&=& e^{-( S_q - S_0)} \text{.}
\end{eqnarray}
At criticality, the usual scaling considerations applied to this object dictate that
\begin{equation}
f(R) \propto \Big( \frac{R}{a} \Big)^{- \Delta_q} \text{,}
\end{equation}
where $\Delta_q$ is the scaling dimension of $m^*_q$ and $a$ is a short-distance cutoff (\emph{e.g.} the lattice spacing).  This implies that
\begin{equation}
S_q - S_0 \sim \Delta_q \ln \Big( \frac{R}{a} \Big) \text{.}
\end{equation}
In the disordered phase ($g > g_c$) there is a finite correlation length $\xi$, and for $R \gg \xi$ one has
\begin{equation}
\label{eqn:action-difference}
S_q - S_0 \sim \Delta_q \ln \Big( \frac{\xi}{a} \Big) \text{.}
\end{equation}
Working in the disordered phase, Murthy and Sachdev directly evaluated $S_q$ and obtained the coefficient of the logarithm in Eq.~(\ref{eqn:action-difference}), and hence the monopole scaling dimension.  In this paper we will calculate the same quantity by a somewhat less direct but technically much simpler method.

As it will be needed later on, we now compute the $N \to \infty$ critical coupling $g_c$, where the phase transition occurs.  On the ${\rm SU}(N)$-symmetric side of the phase diagram, the lowest action saddle point is expected to be given by $A_{\mu} = A^0_{\mu} = 0$ and $i \lambda = i \bar{\lambda}_0 =  m^2$. Thus, the gap equation $\frac{\delta S}{\delta \lambda} = 0$ becomes,
\beq \label{Saddleunreg} \int \frac{d^3 p}{(2 \pi)^3} \frac{1}{p^2 + m^2} = \frac{1}{N g}\eeq
The integral on the left hand side is ultraviolet-divergent and needs to be regularized. We will consistently use throughout this paper Pauli-Villars regularization, which is obtained by augmenting the operator trace in Eq.~(\ref{Snonloc}) by
\begin{equation}   \Tr \ln(-D_{\mu} D_{\mu} + i \lambda) \to 
    \Tr \ln(-D_{\mu}D_{\mu} + i \lambda) + \sum_i s_i \Tr \ln(-D_{\mu} D_{\mu} + i \lambda + M^2_i) \text{,}
\end{equation}
where $M^2_i$ are regulator masses to be taken to infinity, and $s_i$ are alternatingly $-1$ for fermionic regulators and $+1$ for bosonic regulators. To regularize the trace completely in the current problem, we actually need three regulator fields ($i = 1,2,3$), satisfying
\beq \sum_i s_i = -1 \quad \text{and} \quad \sum_i s_i M^2_i = 0 \text{.} \label{regcond}\eeq

Thus, the regularized saddle point equation (\ref{Saddleunreg}) is
\beq \int \frac{d^3 p}{(2 \pi)^3} \left( \frac{1}{p^2 + m^2} + \sum_i s_i \frac{1}{p^2 + m^2 + M^2_i}\right) = \frac{1}{N g} \text{.}\eeq
At the critical point, the $z$-boson mass $m$ vanishes, thus the critical coupling $g_c$ is given by
\beq \int \frac{d^3 p}{(2 \pi)^3} \left(\frac{1}{p^2} + \sum_i s_i \frac{1}{p^2 + M^2_i}\right) = \frac{1}{N g_c} \text{.}
\eeq
Evaluating the integrals, the result is
\beq
 \label{gc} 
 \frac{1}{N g_c} = -\frac{1}{4 \pi} \sum_i s_i M_i \text{.}
\eeq

\section{State-operator correspondence and monopole scaling dimensions}
\label{sec:soc}

While the state-operator correspondence is a standard and well-known feature of conformal field theory (CFT),\cite{yellowbook} it has not been widely applied in condensed matter physics except in the context $D = 2$ CFTs.\endnote{One exception is Ref.~\onlinecite{hermele04}, which used the results of Ref.~\onlinecite{borokhov02} to study the stability of algebraic spin liquids.}   For this reason, in this section we introduce in some detail the state-operator correspondence for a CFT in general space-time dimension $D$.

We consider a CFT in Euclidean space-time invariant under the Euclidean Poincar\'e group and under scale transformations.  (We actually do not need invariance under special conformal transformations for the following discussion.)  We shall work in the scaling limit (\emph{i.e.} continuum limit), so that, in particular, we can think of scale transformations as an exact symmetry.  By assumption, any local operator can be written as a linear combination of scaling operators ${\cal O}_i(x)$.  Scale invariance is the statement that any correlation function of local operators is unchanged upon replacing ${\cal O}_i(x)$ by ${\cal O}'_i(x) = \lambda^{\Delta_i} {\cal O}_i(\lambda x)$, where $\Delta_i$ is the scaling dimension of ${\cal O}_i$.  The Noether current associated with scale transformations is denoted $j^D_{\mu}$.

The goal of the ensuing discussion is twofold.  First, we shall show that there is a quantum Hamiltonian $\hat{H}_S(R)$ defined on the $(D-1)$-sphere of radius $R$.  The eigenstates of this Hamiltonian are in one-to-one correspondence with the scaling operators ${\cal O}_i$, and their energies are related to the scaling dimensions by $E_i = \Delta_i / R$.  Second, we will give a simple method for constructing $\hat{H}_S(R)$.

We shall define the ``spherical Hamiltonian'' $H_S(R)$ on a sphere of radius $R$ centered at the origin:
\begin{equation}
\label{eqn:spherical-hamiltonian}
H_S(R) \equiv \frac{1}{R} \int d^D x \, \delta(|x| - R) \, n_{\mu} j^D_{\mu} \text{.}
\end{equation}
Note that $H_S(R)$ is not quite the same as the quantum Hamiltonian $\hat{H}_S(R)$, which has not yet been defined.  In Eq.~(\ref{eqn:spherical-hamiltonian}), $n_{\mu}(x)$ is the outward normal vector of the sphere, and the initial factor of $1/R$ has been inserted for later convenience.  The spherical Hamiltonian is useful because it is the generator of infinitesimal scale transformations.  This statement is made precise by the Ward identity, which for the scaling operator ${\cal O}_i(x)$ can be written
\begin{equation}
H_S(R) {\cal O}_i(x) =  \frac{1}{R} ( \Delta_i + x_{\mu} \partial_{\mu} ) {\cal O}_i(x) \text{,}
\end{equation}
provided $|x| < R$.  (For a development of Ward identities as they are used here, we refer the reader to Chapter 2 of Ref.~\onlinecite{polchinski-book}.)

We need to construct the Hilbert space in which $\hat{H}_S(R)$ acts.  Suppose the Lagrangian depends on the set of fields $\phi_a$.  A wavefunction on the $(D-1)$-sphere of radius $R$ is a functional
$\Psi = \Psi[ \phi_a]$, which depends only on $\phi_a(x)$ for $|x| = R$.  The operator $\hat{H}_S(R)$ is defined by its action on the wavefunction $\Psi$:
\begin{equation}
[ \hat{H}_S(R) \Psi] [\phi_a] = \lim_{\epsilon \to 0^+}
\int \prod_{R-\epsilon \leq |x| < R + \epsilon} [ d\phi'_a(x) ] \Big[ \prod_{|x| = R + \epsilon} \delta(\phi_a(x) - \phi'_a(x) ) \Big] H_S(R) \, \Psi[\phi_a' ; R - \epsilon] \text{.}
\end{equation}

For each scaling operator, we can associate a wavefunction $\Psi_i$ by inserting ${\cal O}_i$ at the origin, and  ``cutting open'' the path integral at $|x| = R$.  This means we integrate over $\phi_a(x)$ for $|x| < R$, with a fixed boundary condition at $|x| = R$.  Formally,
\begin{equation}
\Psi_i [ \phi_a ; R ] = \int \prod_{|x| < R} [d \phi'_a(x)]  \Big[ \prod_{|x| = R} \delta(\phi_a(x) - \phi'_a(x)) \Big]
{\cal O}_i(0) e^{- S[\phi'_a] } \text{.}
\end{equation}
The action of $\hat{H}_S(R)$ on $\Psi_i$ can be calculated using the Ward identity:
\begin{eqnarray}
[\hat{H}_S(R) \Psi_i][\phi_a] &=&
\lim_{\epsilon \to 0^+} \int \prod_{|x| < R + \epsilon} [ d\phi'_a(x) ] \Big[ \prod_{|x| = R + \epsilon} \delta(\phi_a(x) - \phi'_a(x) ) \Big] H_S(R) {\cal O}_i(0) e^{-S[\phi'_a]} \\
&=&
\frac{\Delta_i}{R}  \int \prod_{|x| < R} [d \phi'_a(x)]  \Big[ \prod_{|x| = R} \delta(\phi_a(x) - \phi'_a(x)) \Big]
{\cal O}_i(0) e^{- S[\phi'_a] } \\
&=& \frac{\Delta_i}{R} \Psi_i[ \phi_a] \text{.}
\end{eqnarray}
Thus we have shown that $\Psi_i$ is an eigenstate of $\hat{H}_S(R)$, where the energy $E_i$ is simply related to the scaling dimension of ${\cal O}_i$ by $E_i = \Delta_i / R$.  Furthermore, this result can be used to argue that for each ${\cal O}_i$ there is a unique state $\Psi_i$.  First, if two ${\cal O}_i$ have different scaling dimensions, then the corresponding states have different energies and are clearly distinct (\emph{i.e.} they are orthogonal).  Suppose that a set of ${\cal O}_i$ have the same scaling dimension.  Generically, this will only occur if these operators form an irreducible multiplet under the global symmetries of the CFT.  The corresponding states must transform under the same multiplet; therefore, they must be linearly independent,  and can be chosen to be mutually orthogonal.

To complete this discussion we still need to show that wavefunctions $\Psi[\phi_a]$ and scaling operators ${\cal O}_i$ are in one-to-one correspondence.  We have already shown that for each scaling operator there is a unique state $\Psi_i$.  It remains to be shown that every eigenstate of $\hat{H}_S(R)$ corresponds to a unique scaling operator.  First, on general grounds of scale invariance, there must be a one-to-one linear mapping relating eigenstates of $\hat{H}_S(R)$ to those of $\hat{H}_S(r)$.  Consider an eigenstate $\Psi[\phi_a ; R]$ of $\hat{H}_S(R)$ with energy $E$, whose image under this mapping is $\Psi[\phi_a; r]$ with energy $E' = E R / r$.  ($E'$ must have this form because the energies scale with inverse radius of the sphere, as is apparent, for example, from the form of the Ward identity.)  We shall be interested in $r < R$, and we may make $r$ as small as we like (as long as it is not so small that we are no longer in the scaling limit).  We consider a functional integral where we insert this state at radius $r$, that is
\begin{equation}
Z_{\Psi} = \int \prod_{r < |x| < \infty} [ d \phi_a(x)] \Psi[\phi_a ; r] e^{- S[\phi_a] } \text{.}
\end{equation}
As $r$ becomes small, we can view this as the insertion of some local operator ${\cal O}$ at the origin.  That is,
\begin{equation}
\lim_{r \to 0} Z_{\Psi}  = \int \prod_{x} [d \phi_a(x)] {\cal O}(0) e^{- S[\phi_a]} \text{.}
\end{equation}
Now we can apply the Ward identity to an insertion of $H_S(R)$ inside $Z_{\Psi}$:
\begin{eqnarray}
 \int \prod_{r < |x| < \infty} [ d \phi_a(x)] \Psi[\phi_a ; r]  H_S(R) e^{- S[\phi_a] } &=&
\frac{r}{R} \lim_{\epsilon \to 0^+} \int \prod_{r < |x| < \infty} [ d \phi_a(x)] \Psi[\phi_a ; r]  H_S(r + \epsilon) e^{- S[\phi_a] }  \\
&=& \frac{E' r}{R}  \int \prod_{r < |x| < \infty} [ d \phi_a(x)] \Psi[\phi_a ; r]  e^{- S[\phi_a] } \\
&=& E  \int \prod_{r < |x| < \infty} [ d \phi_a(x)] \Psi[\phi_a ; r]  e^{- S[\phi_a] } \text{.}
\end{eqnarray}
Taking the limit $r \to 0$, the above relations imply the operator equation $H_S(R) {\cal O}(0) = E {\cal O}(0)$, and ${\cal O}$ is a scaling operator, as desired.

Now that we have established the basic facts of the state-operator correspondence, we will outline a simple procedure to actually construct $\hat{H}_S(R)$.  It is useful to recall how this can be done for the usual Hamiltonian.  Starting from a quantum state defined on the space-like hypersurface at constant imaginary time $\tau$, the Hamiltonian, which generates time translations, can be defined in terms of the transfer matrix $e^{- \delta \tau \hat{H}}$ that evolves to the hypersurface at $\tau + \delta \tau$.  In principle, the transfer matrix can be obtained from the functional integral by integrating over the fields between $\tau$ and $\tau + \delta \tau$.  

Similarly, in the present case we can start with a quantum state defined on the $(D-1)$-sphere of radius $R$.  It is useful to work in polar coordinates  $x = (r, \Omega)$, where $\Omega$ includes the $D - 1$ angular coordinates, and make the change of variables $r = R e^{\tau / R}$ for a fixed value of $R$.  In these variables, scale transformations are realized as ``time'' translations $\tau \to \tau + \delta \tau$.  An infinitesimal scale transformation sends $R \to R e^{\delta \tau / R} = R + \delta\tau$.  Therefore the spherical Hamiltonian, which generates scale transformations, can be obtained from the transfer matrix $e^{- \delta\tau \hat{H}_S(R)}$ that evolves the state at $R$ to one at radius $R + \delta\tau$.

\begin{figure}
\includegraphics[width=4in]{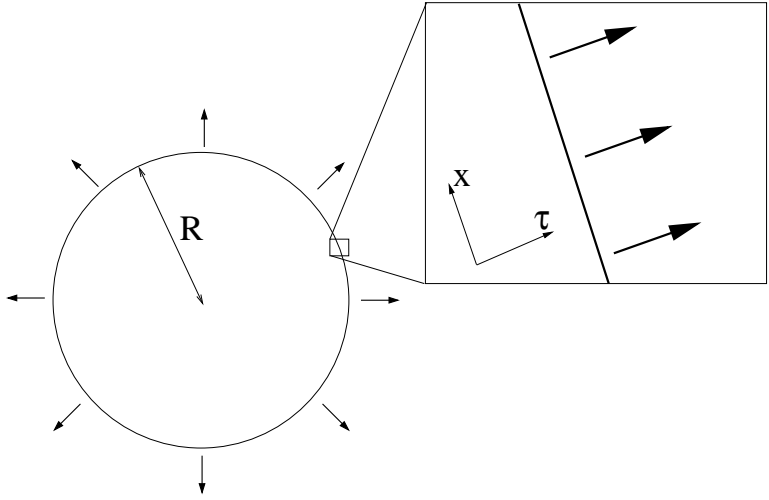}
\caption{Depiction of a scale transformation as an evolution from the sphere at radius $R$ to an expanded sphere with larger radius.  In the magnified region, we illustrate that this evolution is locally equivalent to a time translation, with the locally defined time ($\tau$) and space ($x$) coordinates shown.}
\label{fig:scale-transformation}
\end{figure}

Now, as illustrated in Fig.~\ref{fig:scale-transformation}, for a small patch of the $(D-1)$ sphere of radius $R$, the infinitesimal scale transformation is indistinguishable from an infinitesimal time translation, in the radial direction.  On this small patch, then, the scale transformation will simply be generated by the Hamiltonian density (for appropriately defined local time and space directions).  In order to obtain the generator of scale transformations for the entire sphere, we simply need to wrap the flat-space Hamiltonian onto the sphere.  In practice, it is often easier to work with the functional integral corresponding to $\hat{H}_S(R)$, which is defined on the space $S^{D-1}(R) \times {\mathbb R}$.  Here $S^{D-1}(R)$ is the $(D-1)$-sphere of radius $R$, and ${\mathbb R}$ is the imaginary time direction parametrized by $\tau$.

\section{Calculation}
\label{sec:calc}

Our objective is to compute the scaling dimension $\Delta_q$ of the monopole operator of charge $q$. Such an operator will create states with flux $2 \pi q$ out of the vacuum. Therefore, by the state-operator correspondence, to find $\Delta_q$ we must tune the theory to the critical coupling $g_c$, compactify the spatial manifold to a two-sphere $S^2$ of radius $R$ and find the energy of the state carrying a flux $2 \pi q$ over the sphere.

As a first step we need to find the saddle point of the theory on a sphere with flux. We expect the saddle point for the gauge field $A_{\mu}$ to be given by a uniform distribution of the flux over the spatial sphere (in particular $A_\tau = 0$). We also expect the Lagrange multiplier $\lambda$ to go to a finite constant, $i \lambda = m^2_q$. Note that even though for an infinite system $i \lambda = m^2 = 0$ at the critical point, finite size effects lead to a non-vanishing $m^2_q \sim O(R^{-2})$ on a sphere of radius $R$. In fact, as we will see shortly, $\sqrt{m^2_q + q/2}$ is just the minimal energy to create a spinon above the state with flux $q$. In particular, for $q = 0$, we expect $m_0 R$ to be the scaling dimension $\Delta_z$ of the operator $z$.\footnote{Because $z$ is not gauge invariant, it only makes sense to talk about its scaling dimension in the $N = \infty$ limit, where gauge fluctuations are completely suppressed.} We know that for $N \to \infty$, this conformal dimension is just the engineering dimension for the field $z$ -- namely $\Delta_z = 1/2$.  We will verify shortly that $m_0 R = 1/2$.

By varying the effective action [\emph{i.e.} the analog of Eq.~(\ref{Snonloc}) on the sphere] with respect to $\lambda$ we obtain the gap equation on a sphere with flux,
\beq
\operatorname{Tr} \Big[ \frac{1}{-D_{\mu} D_{\mu} + m^2_q}  \Big] + \sum_i s_i \operatorname{Tr} \Big[ \frac{1}{-D_{\mu} D_{\mu} + m^2_q + M^2_i} \Big] = \frac{4 \pi R^2 \beta}{N g_c} \text{.} \eeq
where $\beta$ is the length of the temporal direction. Using translational invariance along the time direction,
\beq \int \frac{d \omega}{2 \pi} \left( \operatorname{Tr}_{\perp} \Big[ \frac{1}{-D_{\perp}^2+ \omega^2 + m^2_q} \Big] + \sum_i s_i \operatorname{Tr}_{\perp} \Big[ \frac{1}{ -D_{\perp}^2 +\omega^2+ m^2_q + M^2_i} \Big] \right) = \frac{4 \pi R^2}{N g_c} \text{,}\label{gapsphere}\eeq
where $-D^2_{\perp}$ is the square of the covariant derivative along spatial directions, and $\operatorname{Tr}_{\perp}$ is the trace over the space of functions on the sphere of radius $R$. We may take the $\omega$-integral, obtaining 
\beq \frac{1}{2} \left(\operatorname{Tr}_{\perp} \Big[ \frac{1}{(-D_{\perp}^2+ m^2_q)^{\frac12}} \Big] + \sum_i s_i \operatorname{Tr}_{\perp} \Big[ \frac{1}{ (-D_{\perp}^2 + m^2_q+ M^2_i)^{\frac12}} \Big] \right) = \frac{4 \pi R^2 }{N g_c} \text{.} \label{gapsphere2}\eeq

To evaluate the traces in (\ref{gapsphere2}) we need the spectrum of $-D^2_{\perp}$. Fortunately, this problem of a particle moving on a sphere with a monopole of charge $q$ at the origin was solved a long time ago by Wu and Yang.\cite{wu76} The eigenfunctions are the monopole harmonics, $Y_{q/2,l,m}$ with $l = q/2,\, q/2+1, ...$ and $m = -l,\, -l +1, ..., l$. The corresponding eigenvalue of $-D^2_{\perp} R^2$ is $l(l+1) - (q/2)^2$. Note that, for $q = 0$, we recover the usual spherical harmonics. 
Thus, Eq.~(\ref{gapsphere2}) becomes
\beq \frac{1}{2} \sum_{l=q/2}^{\infty}  \frac{2 l + 1}{4 \pi R} \left(\frac{1}{(l(l+1)-(q/2)^2+ (m_q R)^2)^{\frac12}} + \sum_i s_i \frac{1}{(l(l+1)-(q/2)^2+ (m_q R)^2 + (M_i R)^2)^{\frac12}}\right) = \frac{1}{N g_c} \text{.}\label{gapsphere3}\eeq

We would like to isolate the cutoff dependence of the left-hand side of Eq.~(\ref{gapsphere3}). For this purpose, we rewrite Eq.~(\ref{gapsphere3}) as
\beq \frac{4 \pi R}{N g_c} = G_q(a^2_q) + \sum_i s_i G_q(b^2_{qi}) \label{gapUVsep} \text{,}\eeq 
where
\bea G_q(b^2) &=& \sum_{l=q/2}^{\infty} \left(\frac{l+1/2}{((l+1/2)^2+b^2)^{\frac12}}-1\right)\label{Gqdef}\\
a_q^2 &=& (m_q R)^2 -\frac{1}{4} (q^2 + 1), 
\quad b^2_{qi} = a^2_q + (M_i R)^2 \text{.} \eea
Here we have used the fact $\sum_i s_i = -1$. 
The ultraviolet cutoffs $M_i$ now appear only in the second term on the right hand side of Eq.~(\ref{gapUVsep}). To finish isolating the cutoff dependence we need to find the behavior of the function $G_q(b^2)$ in the limit $b^2 \to \infty$. This is easily accomplished using Poisson resummation (see Appendix~\ref{app:functions}), and we obtain
\beq G_q(b^2) \sim -b + q/2\label{Gbfinal}, \quad b \to \infty \text{.}\eeq
Here, we have dropped terms decaying as $b^{-1}$ or faster. Substituting this into Eq.~(\ref{gapUVsep}),
we have
\beq \frac{4 \pi R}{N g_c} =  G_q(a^2_q) - \sum_i s_i b_{qi} - q/2\eeq
Now, eliminating $g_c$ using equation Eq.~(\ref{gc}) , we see that the ultraviolet-divergent terms cancel, and we obtain
\beq G_q(a^2_q) = q/2 \text{.} \label{GapFinal}\eeq
This is precisely Eq.~(3.23) of Murthy and Sachdev,\cite{murthy90} with the identification $\alpha_q =  -a^2_q - q^2/4 = 1/4 - (m_q R)^2$. Also, notice that $G_q(0) = 0$. So for $q = 0$, we immediately obtain $a^2_0 = 0$ as  the solution  to Eq.~(\ref{GapFinal}), and $m_0 R = 1/2$ as expected.

Now we proceed to the calculation of the energy of a state with flux $2 \pi q$. Namely, let
\beq T_q = \frac{1}{\beta} \left(\Tr \ln(-D_{\mu} D_{\mu} + m_q^2) + \sum_i s_i \Tr \ln(-D_{\mu} D_{\mu} + m^2_q + M^2_i)\right) \text{.} \eeq
The saddle-point action of the configuration with flux $2 \pi q$ is given by 
\beq S_q = \beta N T_q - \frac{1}{g_c} \int dx \, m^2_q = N \beta (T_q  - \frac{4 \pi R^2}{N g_c} m^2_q) \text{,}\eeq
and, therefore, the energy $E_q$ is given by
\beq \frac{E_q R}{N} = T_q R - \frac{4 \pi R}{N g_c} (m_q R)^2 \label{ET} \text{.} \eeq
The desired scaling dimension of the charge-$q$ monopole operator is $\Delta_q = (E_q - E_0) R$. 

We now evaluate $T_q$. Going to frequency space, we have
\bea T_q &=& \int \frac{d \omega}{2 \pi} \left(\Tr_{\perp} \ln(-D^2_\perp +\omega^2+ m_q^2) + \sum_i s_i \Tr_{\perp} \ln(-D^2_\perp +\omega^2+ m^2_q + M^2_i)\right) \\
&=& \Tr_{\perp} (-D^2_\perp + m_q^2)^{\frac12} + \sum_i s_i \Tr_{\perp} (-D^2_\perp + m_q^2+M^2_i)^{\frac12} \text{.}\eea
Recalling the form of the spectrum of $-D^2_\perp$,
\beq T_q R = \sum_l (2 l + 1) \left( (l (l+1) - (q/2)^2 + (m_q R)^2)^\frac12 + \sum_i s_i (l (l+1) - (q/2)^2 + (m_q R)^2 + (M_i R)^2)^\frac12\right) \text{.} \eeq
We rewrite this in the form
\beq T_q R = 2 F_q(a^2_q) + 2 \sum_i s_i F_q(b^2_{qi})\label{TF} \text{,}\eeq
where
\beq F_q(b^2) = \sum_{l=q/2}^{\infty} \left((l+1/2) ((l+1/2)^2+b^2)^{\frac12} - (l+1/2)^2 - \frac{1}{2} b^2\right)\label{Fq} \text{.}\eeq 
It should be noted that the sum over $l$ in Eq.~(\ref{Fq}) converges. As in the analysis of the gap equation, only the second term of Eq.~(\ref{TF}) depends on the ultraviolet cutoff. Also as before, we consider the $b \to \infty$ limit of $F_q(b^2)$. After a short calculation (see Appendix~\ref{app:functions}), we obtain
\beq F_q(b^2) \sim -\frac{1}{3} b^3 + \frac{q}{4} b^2 + (\frac{1}{24} - \frac{q^2}{8}) b + \frac{1}{24} q (q^2-1), \quad b \to \infty\label{Fqeven} \text{.} \eeq
Substituting this result into Eq.~(\ref{TF}) and noting that
$b^3_{qi} = (M_i R)^3 + \frac{3}{2} a^2_q (M_i R) + {\cal O}[(M_i R)^{-1}]$, we find
\beq T_q R = -\frac{2}{3} \sum_i s_i (M_i R)^3 + \left(\frac{1}{12} - \frac{q^2}{4} - a^2_q\right) \sum_i s_i  M_i R + 2 F_q(a^2_q) - \frac{q}{2} a^2_q
-\frac{1}{12} q(q^2-1) \label{Tfinal} \text{.} \eeq

Now, we can bring everything together.  Substituting the critical coupling $g_c$ [Eq.~(\ref{gc})] into Eq.~(\ref{ET}) and
recalling that $(m_q R)^2 = a^2_q + \frac{1}{4} (q^2 + 1)$, we find
\beq \frac{E_q R}{N} = -\frac{2}{3} \sum_i s_i (M_i R)^3 + \frac{1}{3} \sum_i s_i (M_i R)+ 2 F_q(a^2_q) - \frac{q}{2} a^2_q -
\frac{1}{12} q (q^2-1)\label{Eqfinal}\eeq
The cutoff-dependent (and also ultraviolet-divergent) terms in $E_q R / N$ comprise a $q$-independent constant. Hence, the energy differences are finite:
\beq \frac{(E_q-E_0)R}{N} = 2 (F_q(a^2_q) - F_0(a^2_0)) - \frac{q}{2} a^2_q - \frac{1}{12} q (q^2-1)\eeq
Recalling that $a^2_0 = 0$ and noting that $F_q(0) = 0$, we obtain the final result,
\beq \label{Deltafin}\frac{\Delta_q}{N} = \frac{(E_q-E_0)R}{N} = 2 F_q(a^2_q) - \frac{q}{2} a^2_q - \frac{1}{12} q (q^2-1) \text{.} \eeq
It is easy to show this result is precisely that of Murthy and Sachdev (see Appendix~\ref{app:sameresult}).

\section{Discussion}
Let us put our calculation into the context of the role of ${\rm U}(1)$ flux symmetry in the noncompact $\cpn$ model. In the ordered phase of the theory ($g<g_c$) the flux symmetry is unbroken, as the Meissner effect leads to flux confinement. The configurations carrying  magnetic flux in this phase have a finite energy and, in fact, are quantum descendants of instantons of the two-dimensional $\cpn$ model.\cite{luscher78} Close to the critical point these instantons are strongly dressed by the interaction: their size grows and their energy decreases as $g \to g_c$. Precisely at the QCP the instantons become massless. The condition that flux and spin gaps vanish at the same critical point is at the heart of deconfined criticality. We have verified this fact explicitly in the present paper by showing that the energy of a flux $q$ instanton goes as $\Delta_q/R$ on a sphere of radius $R$.  The observation that on a finite sphere the energy scales as $1/R$ at the QCP follows from dimensional analysis arguments. However, the fact that $\Delta_q$ coincides with the scaling dimension of the monopole operator is a non-trivial prediction of the state-operator correspondence of conformal field theory. The agreement between our result and the more direct computation of $\Delta_q$ by Murthy and Sachdev is a strong check that the monopole operator survives in the scaling limit. 

Now, to complete our discussion, once the coupling $g > g_c$ and we are in the disordered phase, the instantons, having become massless at the phase transition, condense. As a result, the ${\rm U}(1)$ flux symmetry is spontaneously broken; the photon is a Goldstone boson associated with this symmetry, since it is created out of the vacuum by the current $j^{G}_{\mu}$. What is the fate of configurations carrying finite flux in this phase? We can compute their energy directly from the effective action for the photon field,
\beq S = \frac{1}{2 e^2} \int d^3 x (\epsilon_{\mu \nu \lambda} \d_{\nu} A_{\lambda})^2 \eeq
where to leading order in $1/N$, $e^2 = 24 \pi m/N$, with $m$  the spinon mass. For simplicity we work with a flat spatial manifold here (e.g. a torus). Then, smearing the flux $2 \pi q$ uniformly over the space, 
\beq \epsilon_{i j} \d_i A_j = \frac{2 \pi q}{V} \eeq
where $V$ is the spatial volume. The energy becomes,
\beq E = \frac{(2 \pi q)^2}{2 e^2 V}\eeq
Indeed, as always occurs when a continuous global symmetry is spontaneously broken, the states of finite charge (flux) form a tower, with energies scaling as inverse volume.

Thus, in the $N = \infty$ limit, we have a detailed quantitative understanding of the flux sector of the $\cpn$ model at the critical point and in the disordered phase. It would be interesting to extend the quantitative description to the ordered phase. In particular, it would be interesting to compute the finite instanton mass, $m_i$, which we expect to govern the long distance decay of monopole-antimonopole correlation functions. From general scaling arguments, we expect $m_i \sim (g-g_c)^{\nu}$, where $\nu$ is the correlation length exponent. Moreover, we expect the ratio $m_i/\rho_s$, where $\rho_s$ is the spin-stiffness, to be a universal number. Unfortunately, it is rather difficult to analyze the instantons in the ordered phase even at $N = \infty$, since the saddle point value of the fields $A_{\mu}$ and $z_{\alpha}$ is no longer dictated by symmetry as it was at the critical point.

\label{sec:discussion}
\section{Conclusion}
\label{sec:conclusion}
In this paper we have used the state-operator correspondence of conformal field theory to compute the monopole scaling dimension in the
$\cpn$ model at $N = \infty$. Our result agrees with the more direct calculation by Murthy and Sachdev;\cite{murthy90} however, our approach has the advantage of technical simplicity. In fact, one can even envision using this method to compute the $1/N$ corrections to the monopole scaling dimension. 
From the conceptual point of view our result demonstrates the vanishing of the flux gap at the QCP and confirms the survival of the monopole operator in the scaling limit.

\acknowledgments{We are grateful to Subir Sachdev and Ariel Zhitnitsky for useful discussions.  This research is supported by NSF Grants No.  DMR-0529399 (M.P.A.F.), DMR-0705255 (T.S.) and DMR-0757145 (M.M.).}
\appendix
\section{Functions $G_q(b^2)$ and $F_q(b^2)$}
\label{app:functions}
The purpose of this appendix is to compute the behaviour of functions $G_q(b^2)$ [Eq.~(\ref{Gqdef})], and $F_q(b^2)$ [Eq.~(\ref{Fq})], in the limit $b \to \infty$. 

We begin with $G_q$. First, we consider the case $q$-even. Then,
\beq G_q(b^2) = \sum_{l=0}^{\infty} \left(\frac{l+1/2}{((l+1/2)^2+b^2)^{\frac12}}-1\right) - \sum_{l=0}^{q/2-1} \left(\frac{l+1/2}{((l+1/2)^2+b^2)^{\frac12}}-1\right)\text{.}\eeq
In what follows, we will drop all the corrections to $G_q(b^2)$ that vanish as $b^{-1}$ or faster. Thus, simplifying the second term above,
\beq G_q(b^2) = \sum_{l=0}^{\infty} \left(\frac{l+1/2}{((l+1/2)^2+b^2)^{\frac12}}-1\right) + q/2 \text{.} \eeq
Now we utilize the symmetry of the summand under $l \to -l -1$, obtaining
\beq G_q(b^2) = \frac{1}{2} \sum_{l=-\infty}^{\infty} \left(\frac{|l+1/2|}{((l+1/2)^2+b^2)^{\frac12}}-1\right) + q/2 \text{.} \eeq
Upon Poisson-resumming the $l$'s, we have
\bea G_q(b^2) &=& \frac{1}{2} \sum_{n = -\infty}^{\infty} (-1)^n \int_{-\infty}^{\infty} dl \left(\frac{|l|}{(l^2+b^2)^{\frac12}}-1\right) e^{2 \pi i n l} + q/2\label{Poiss}\\
&=& \int_{0}^{\infty} dl \left(\frac{l}{(l^2+b^2)^{\frac12}}-1\right) + 2 \sum_{n=1}^{\infty} (-1)^n \int_0^{\infty} dl \left(\frac{l}{(l^2+b^2)^{\frac12}}-1\right)\cos(2\pi n l) + q/2 \text{.}\nn\eea
As usual, the leading (divergent) contribution in the $b \to \infty$ limit comes from the $n = 0$ term in Eq~(\ref{Poiss}), which is
\beq \int_{0}^{\infty} dl \left(\frac{l}{(l^2+b^2)^{\frac12}}-1\right) = -b \text{.} \eeq
As for the $n \ge 1$ terms, we rotate the contour of integration as follows:
 \bea & &\int_0^{\infty} dl \left(\frac{l}{(l^2+b^2)^{\frac12}}-1\right)\cos(2\pi n l) = b \, Re \int_0^{\infty} dl \left(\frac{l}{(l^2+1)^{\frac12}}-1\right)e^{2 \pi i n b l}\\
 &=& b\, Re \int_0^{\infty} i dy \left( i y (\theta(1-y) (1-y^2)^{-\frac12} + \theta(y-1) (y^2-1)^{-\frac12} e^{-i \pi/2}) - 1\right)e^{-2 \pi n b y}\nn\\ &=& -b \int_0^{1} dy \,\frac{y}{(1-y^2)^{\frac12}} e^{-2 \pi n b y} = -b \left(\frac{1}{(2 \pi n b)^2} + O\left(\frac{1}{(n b)^{4}}\right)\right) \text{.}\eea 
Hence,
\beq 2 \sum_{n=1}^{\infty} (-1)^n \int_0^{\infty} dl \left(\frac{l}{(l^2+b^2)^{\frac12}}-1\right)\cos(2\pi n l) \sim \frac{2}{(2 \pi)^2 b} \sum_{n=1}^{\infty} (-1)^{n+1} \frac{1}{n^2} \sim O(b^{-1}) \text{,} \eeq
and this can be dropped in the limit $b \to \infty$. Therefore,
\beq G_q(b^2) \sim -b + q/2\label{Gbfinala}\eeq

Repeating this analysis for $q$-odd and $b \to \infty$,
\bea G_q(b^2) &=& \sum_{l= \frac{q+1}{2}}^{\infty} \left(\frac{l}{(l^2+b^2)^{\frac12}} - 1\right) = \sum_{l=0}^{\infty}\left(\frac{l}{(l^2+b^2)^{\frac12}} - 1\right) - \sum_{l=0}^{\frac{q-1}{2}} \left(\frac{l}{(l^2+b^2)^{\frac12}} - 1\right)\\
&=& \frac{1}{2} \sum_{l=-\infty}^{\infty}\left(\frac{|l|}{(l^2+b^2)^{\frac12}} - 1\right) - 1/2 + \frac{q+1}{2} = \frac{1}{2} \sum_{l=-\infty}^{\infty}\left(\frac{|l|}{(l^2+b^2)^{\frac12}} - 1\right) + q/2\nn\\
&=& \frac{1}{2} \sum_{n=-\infty}^{\infty} \int dl \,\left(\frac{|l|}{(l^2+b^2)^{\frac12}} - 1\right) e^{2 \pi i n l} + q/2 \text{.} \label{Poissqodd}\eea
Comparing Eq.~(\ref{Poissqodd}) to its $q$-even counterpart Eq.~(\ref{Poiss}), we see that the  only difference is the absence of the factor $(-1)^n$ in the sum. Recalling that in the $b \to \infty$ limit the only finite contribution came from the $n=0$ term in the sum, we obtain the same result as in the $q$-even case [Eq.~(\ref{Gbfinala})].

Now, we proceed to the function $F_q$. We again begin with the case of $q$ even:
\bea F_q(b^2) &=& \sum_{l=0}^{\infty} \left((l+1/2) ((l+1/2)^2+b^2)^{\frac12} - (l+1/2)^2 - \frac{1}{2} b^2\right)\\  &-& \sum_{l = 0}^{q/2-1} \left((l+1/2) ((l+1/2)^2+b^2)^{\frac12} - (l+1/2)^2 - \frac{1}{2} b^2\right) \text{.}\eea
As before, we drop all the terms decaying as $b^{-1}$ or faster. Thus,
\bea F_q(b^2) &\sim& \frac{1}{2} \sum_{l= -\infty}^{\infty} \left(|l+1/2| ((l+1/2)^2+b^2)^{\frac12} - (l+1/2)^2 - \frac{1}{2} b^2\right)\\ &+& \frac{1}{2} b^2 \sum_{l=0}^{q/2-1} 1 - b \sum_{l = 0}^{q/2-1} (l+1/2) + \sum_{l=0}^{q/2-1} (l+1/2)^2 \text{.}\eea
Poisson-resumming the first sum, we have
\bea F_q(b^2) &\sim& \frac{1}{2} \sum_{n=-\infty}^{\infty} (-1)^n \int_{-\infty}^{\infty} dl \, \left(|l|(l^2+b^2)^{\frac12} - l^2 - \frac{1}{2} b^2\right)e^{2 \pi i n l} + \frac{q}{4} b^2 - \frac{q^2}{8} b + \frac{1}{24} q (q^2-1)\nn\\
&=& \int_0^{\infty} dl \, \left( l (l^2+b^2)^{\frac12} - l^2 - \frac{1}{2} b^2\right) + 2 \sum_{n=1}^{\infty} (-1)^n \int_0^{\infty} dl \left( l (l^2+b^2)^{\frac12} - l^2 - \frac{1}{2} b^2\right) \cos(2 \pi n l) \nn\\&+& \frac{q}{4} b^2 - \frac{q^2}{8} b + \frac{1}{24} q (q^2-1) \text{.}\eea
As before, the most divergent piece in the $b \to \infty$ limit comes from the $n = 0$ term in the Poisson-resummed series, which is
\beq \int_0^{\infty} dl \, \left( l (l^2+b^2)^{\frac12} - l^2 - \frac{1}{2} b^2\right) = - \frac{1}{3} b^3 \text{.}\eeq
The integral for the $n \neq 0$ terms can again be analyzed by rotating the integration contour:
\bea  &&\int_0^{\infty} dl \left( l (l^2+b^2)^{\frac12} - l^2 - \frac{1}{2} b^2\right) \cos(2 \pi n l) =  b^3 \, Re \int_0^{\infty} dl \left( l (l^2+1)^{\frac12} - l^2 - \frac{1}{2} \right) e^{2 \pi i n b l} \nn\\ &=& b^3 Re \int_0^{\infty} i \,dy  \left(i y ( \theta(1-y) (1-y^2)^{\frac12} + \theta(y-1)
(y^2-1)^{\frac12} e^{i \pi/2}) + y^2 - \frac{1}{2}\right) e^{-2 \pi n b y}\nn\\&=& - b^3 \int_0^{1} dy\, y (1-y^2)^{\frac12} e^{-2 \pi n b y} \to
- b^3 \left(\frac{1}{(2 \pi n b)^2} + O(\frac{1}{(nb)^4})\right) \text{.} \eea
Here, unlike for the gap equation, we cannot limit ourselves to just the $n=0$ term in the $b \to \infty$ limit, and
\bea F_q(b^2) &\sim& -\frac{1}{3} b^3 + \frac{2}{(2 \pi)^2} b \sum_{n = 1}^{\infty} \frac{(-1)^{n+1}}{n^2} + \frac{q}{4} b^2 - \frac{q^2}{8} b + \frac{1}{24} q (q^2-1)\\
&=& -\frac{1}{3} b^3 + \frac{q}{4} b^2 + (\frac{1}{24} - \frac{q^2}{8}) b + \frac{1}{24} q (q^2-1) \text{.}\label{Fqevenap}\eea

Performing a similar analysis for $q$-odd, we find
\bea F_q(b^2) &=& \sum_{l=0}^{\infty} \left(l(l^2+b^2)^{\frac12} - l^2 - \frac{1}{2} b^2\right)- \sum_{l = 0}^{(q-1)/2} \left(l (l^2+b^2)^{\frac12} - l^2 - \frac{1}{2} b^2\right)
\\ &\sim& \frac{1}{2} \sum_{l = -\infty}^{\infty} \left(l(l^2+b^2)^{\frac12} - l^2 - \frac{1}{2} b^2\right) - \frac{1}{4} b^2 + \frac{1}{2} b^2 \sum_{l=0}^{(q-1)/2} 1 -b \sum_{l=0}^{(q-1)/2} l + \sum_{l=0}^{(q-1)/2} l^2 \nn\\
&=& \frac{1}{2} \sum_{n=- \infty}^{\infty} \int dl \left(l(l^2+b^2)^{\frac12} - l^2 - \frac{1}{2} b^2\right) e^{2 \pi i n l} + \frac{q}{4}
b^2 -\frac{1}{8} (q^2-1) b + \frac{1}{24} q (q^2 -1)\nn\\
&\sim& -\frac{1}{3} b^3 - \frac{2}{(2 \pi)^2}b \sum_{n=1}^{\infty} \frac{1}{n^2}+\frac{q}{4}
b^2 -\frac{1}{8} (q^2-1) b + \frac{1}{24} q (q^2 -1)\\
&=&  -\frac{1}{3} b^3 +\frac{q}{4}
b^2 +(\frac{1}{24}-\frac{q^2}{8})  b + \frac{1}{24} q (q^2 -1) \text{.} \eea
which is equal to the result [Eq.~(\ref{Fqevenap})] we obtained for $q$-even.

\section{Comparison to Murthy-Sachdev expression}
\label{app:sameresult}
Murthy and Sachdev\cite{murthy90} have expressed their result for the scaling dimension of the monopole operator as
\beq \frac{\Delta_q}{N} = - \Omega_q + \Xi_q + \frac{q^3}{24} + \frac{q}{12} \text{,}\eeq
where
\beq \Omega_q = \frac{q^4}{4}\sum_{l = q/2}^{\infty} \left(\frac{1}{(\sqrt{(2l+1)^2-q^2} + 2 l + 1)^2}\right) \text{,}\eeq
and
\beq \Xi_q =  -\sum_{l=q/2}^{\infty} (2 l + 1) \left(((l+1/2)^2-q^2/4)^{\frac12} - 
((l+1/2)^2-q^2/4- \alpha_q)^{\frac12} - \frac{\alpha_q}{2 ((l+1/2)^2-q^2/4-\alpha_q)^{\frac12}}\right)\label{Sigma} \text{.}\eeq
Using the identification $\alpha_q = -q^2/4 - a_q^2$ to convert this to the notation used in our analysis, and summing the last term in Eq.~(\ref{Sigma}) 
using the gap equation Eq.~(\ref{GapFinal}), we have
\beq \label{Sigmamod} \Xi_q = 2 \sum_{l=q/2}^{\infty} \left((l+1/2)\big(((l+1/2)^2+a^2_q)^{\frac12} - ((l+1/2)^2-q^2/4)^{\frac12}\big) + \frac{\alpha_q}{2}\right) + \frac{q}{2} \alpha_q \text{.}\eeq
For $\Omega_q$, we can eliminate the irrationality in the denominator to obtain
\beq \label{Omegamod} \Omega_q = -2 \sum_{l=q/2}^{\infty} \left((l+1/2)((l+1/2)^2-q^2/4)^{\frac12} - (l+1/2)^2 + \frac{1}{8}q^2\right) \text{.} \eeq
Thus, adding Eq.~(\ref{Sigmamod}) and Eq.~(\ref{Omegamod}),
\beq \frac{\Delta_q}{N} = 2 \sum_{l=q/2}^{\infty} \left((l+1/2)((l+1/2)^2 +a^2_q)^{\frac12} - (l+1/2)^2 - \frac{1}{2} a^2_q\right) -\frac{q}{2} a^2_q - \frac{1}{12} q(q^2-1) \eeq
which is identical to our result [Eq.~(\ref{Deltafin})]. 

\bibliography{msd_paper2}



\end{document}